\documentstyle[12pt]{article}
\begin{document}
\begin{center}
{\bf BOSONISATION EXCERCISE IN THREE DIMENSIONS:\\
GAUGED MASSIVE THIRRING MODEL\\}
\vskip 1cm
Subir Ghosh\footnote{e-mail address: subir@bose.ernet.in}\\
Physics Department,\\
Dinabandhu Andrews College, Calcutta 700084,\\
India.\\
\end{center}
\vskip 2cm
{\bf Abstract}
\noindent

Bosonisation of the massive Thirring model, with a non-minimal
and non-abelian gauging is studied in 2+1-dimensions. The static 
abelian model is solved completely in the large fermion mass limit and
the spectrum is obtained. The non-abelian model is solved for a restricted
class of gauge fields. In both cases explicit expressions for bosonic
currents corresponding to the fermion currents are given.

\newpage
\noindent {\bf Introduction}
\vskip .5cm
\noindent
Ever since the explicit demonstration in 1+1-dimensions of the equivalence
of the massive Thirring model and the sine-Gordon model order by order
in perturbation theory in the charge zero sector \cite{col} and the
subsequent construction of the fermion operator by boson operators
\cite{man}, the concept of bosonisation has proved to be an extremely
useful one. However it was thought that this equivalence is an exclusive
property of one dimensional space, where in reality there is no spin to
distinguish fermions from bosons. Indeed, attempts to generalise 
bosonisation in two space dimensions met with limited success \cite{lush}.

Renewed interest in 2+1-dimensional bosonisation has created a flurry
of activity in recent years, where the problem is attacked from a
different angle. The nonlocal fermion 
determinant generates local terms in the one loop 
 perturbative evaluation, in the limit of large fermion
mass. In the lowest orders of 
inverse fermion mass, the bosonised theory of the 
(abelian) massive Thirring model
turns out to be 
Maxwell-Chern-Simons \cite{fs}. In fact the equivalence between massive
Thirring and $CP(1)$ models in the large fermion mass limit was established
much ago \cite{dr}.
 The situation is not that clear
in the nonabelian models. For example, the $SU(2)$ Thirring model, in the
limit where the Thirring coupling vanishes, can be identified with the 
$SU(2)$ Yang-Mills-Chern-Simons theory, in the limit where the Yang-Mills
term vanishes \cite{bbg}.

In the present work, we consider a theory
of non-minimally gauged Dirac fermions, with a Thirring 
\cite{th} current-current self
interaction. Both abelian and non-abelian gauge groups have been
investigated.
The model lives in 2+1-dimensions.
We study the one loop bosonised version 
of the model in the large fermion mass ($m$)
limit and keep only Chern-Simons ($m$ independent)
and Yang-Mills or Maxwell (O($m^{-1}$) terms. 
 The effect of still higher order terms in the inverse
fermion mass is qualitatively discussed in the abelian context.
 The mapping between the 
fermion and the boson fields at the 
level of currents is obtained. The behaviour of the bosonic charge
operator is studied in detail.

The paper is organised as follows:
Section {\bf II} deals with the nonabelian fermion model 
and its bosonisation. 
In Section {\bf III} we discuss in detail the abelian theory.
Section {\bf IV} contains results for the nonabelian theory for a
special class of gauge fields.
The paper ends with a brief conclusion in Section {\bf V}.
\vskip 1cm

\noindent {\bf II. Bosonisation}
\vskip 0.5cm
The parent non-abelian
fermion model that we wish to study is a system of 
 Dirac fermions with a non-minimal gauging. There is a 
Thirring \cite{th} type current-current self
interaction term as well. The Lagrangian considered by us is
\begin{equation}
L=\bar\psi i\gamma^\mu D_\mu\psi-m\bar\psi\psi
+{g\over 2}(\bar\psi\gamma^\mu T^a\psi)^2.
\label{eqlf}
\end{equation}
The covariant derivative is defined as
$$D_\mu=\partial_\mu-i\gamma A_\mu^aT^a-i\sigma K_\mu^aT^a\equiv\partial_\mu
-i\bar{A}_\mu^aT^a,~~
K_\mu=\epsilon_{\mu\nu\lambda}A^{\nu\lambda}.$$
The anti-hermitian generators satisfy $[T^a,T^b]=f^{abc}T^c $ and
the $\gamma$-matrices are defined via the Pauli matrices by
$$\gamma^0=\sigma^3,~~\gamma^1=i\sigma^1,~~\gamma^2=i\sigma^2.$$
To keep track of the various combinations of vector fields that will
appear, we introduce the notations,
$$(D_\mu)^{(W)ab}=\partial_\mu\delta^{ab}-\rho f^{abc}W_\mu^c;~~
W_{\mu\nu}^a=\partial_\mu W_\nu^a-\partial_\nu W_\mu^a
+\rho W_\mu^b W_\nu^c,$$
where $W$ is some arbitrary vector field and $\rho$ the associated
coupling constant. The fermion current enjoys a conservation law,
\begin{equation}
(D_\mu^{\bar A}J^\mu)^a=o;~~~{\bar A}_\mu=\gamma A_\mu+\sigma K_\mu.
\label{eqferc}
\end{equation}

Note that no gauge field kinetic term 
such as the Yang-Mills or Chern-Simons term, 
is kept in the fermion model as they
will be generated in the bosonisation process, along with other mixed terms.
Hence even
if such terms are kept, their
coefficients will get renormalised by bosonisation.

The usual scheme of linearising the Thirring term in (\ref {eqlf}) is
by introducing an auxiliary field $B_{\mu}^a $, such that when $B_{\mu}^a $
is integrated out, the original model is reproduced. This gives us
$$
\bar L=\bar\psi\gamma^\mu (iD_\mu+B_\mu)\psi-m\bar\psi\psi
-{1\over {2g}}B_\mu^aB^{\mu a}$$
\begin{equation}
=\bar\psi \gamma^\mu(i\partial_\mu+C_\mu)\psi-m\bar\psi\psi
-{1\over {2g}}B_\mu^aB^{\mu a},
\label{eqlb}
\end{equation}
where $C_\mu\equiv B_\mu+\gamma A_\mu+\sigma K_\mu$. The quadratic term in
$B_\mu$ constitutes a mass term for $B_\mu$.
This leads us to the evaluation of the fermion determinant, which is
in general non-local, but yields local expressions under various
approximation schemes. A gauge invariant Pauli-Villars regularisation
has been invoked.
We choose, in particular, the large fermion mass
limit such that $m^{-1}$ is a small term. This also restricts
us to the low energy or long wavelength limit, where terms with smaller
number of derivatives dominate. The Seeley coefficients in the fermion
determinant are computed at the one loop level. With these restrictions,
the bosonised Lagrangian is the following:
\begin{equation}
L_B=-{a\over 4}C_{\mu\nu}^aC^{\mu\nu a}-{1\over {2g}}B_\mu^aB^{\mu a}
+\alpha\epsilon^{\mu\nu\lambda}C_\mu^a(\partial_\nu C_\lambda^a+{1\over 3}
f^{abc}C_\nu^bC_\lambda^c).
\label{eqlbos}
\end{equation}
The coefficients $\alpha=1/(4\pi)$ and $a=-1/(24\pi m)$ are 
known from bosonization rules.

Since there are a number of fields, coupling constants and parameters, a
glossary of the dimensions of them in the $c~=~\hbar~=~1$ system 
of units is
provided below, with $l$ denoting length,
$$[C_\mu]=[B_\mu]=[\psi]=[m]={1\over l},~~[A_\mu]=[\gamma]={1\over{\sqrt l}},
~~[g]=[a]=l,~~[\sigma]=\sqrt l.$$
The
Lagrangian equations of motion following from (\ref{eqlbos}) are,
\begin{equation}
2\sigma\epsilon^{\nu\mu\lambda}(D_\mu^{(A)}X_\lambda)^a+\gamma X^{\nu a}=0,
\label{eqA}
\end{equation}
where
$$
X^{\nu a}=(aD_\mu^{(C)}C^{\mu\nu}
+\alpha\epsilon^{\nu\mu\lambda}C_{\mu\lambda})^a.$$
\begin{equation}
X_\nu^a-{1\over g}B_\nu^a=0.
\label{eqB}
\end{equation}
Putting (\ref{eqB}) in (\ref{eqA}) we get
\begin{equation}
2\sigma\epsilon^{\nu\mu\lambda}(D_\mu^{(A)}B_\lambda)^a+\gamma B^{\nu a}=0,
\label{eqAB}
\end{equation}
This is essentially a generalised nonabelian self dual equation for
$B_\mu^a$.
Our next task is to identify the operator that will correspond to the
fermion current $J_{\mu}^a=\bar{\psi}\gamma_{\mu}T^a\psi $. 
The standard procedure is to introduce a source term 
$\sigma_{\mu}^a J_{\mu}^a $ in the fermion Lagrangian where
$\sigma_{\mu}^a$ is an auxiliary field coupled to the operator $J_\mu^a$
in question. After bosonising this modified Lagrangian, ${{\delta L_B}\over
{\delta\sigma_{\mu}^a}}\mid_{\sigma=0}$ can be identified as the mapping
of the fermion current. This shows us that the bosonised 
current $j^a_{\mu}$ is,
\begin{equation}
j^{a\nu}\equiv X^{\nu a}=(aD_\mu^{(C)}C^{\mu\nu}
+\alpha\epsilon^{\nu\mu\lambda}C_{\mu\lambda})^a
={1\over g}B_\nu^a.
\label{eqbosc}
\end{equation}
The last equality follows from (\ref{eqB}).
It is reassuring to note that the whole structure is internally consistent
since in the fermion model, the equation of motion for $B_{\mu}^a$ is
$${1\over g}B_{\mu}^a=J_{\mu}^a.$$
The above operator identity is preserved now as well,
$${1\over g}B_{\mu}^a=j_{\mu}^a.$$
The fermion current conservation equation in (\ref{eqferc}) in abelian theory
reduces to
$$\partial_\mu J^\mu=0.$$
From the expression of $X_{\mu}^a$ or from the nonabelian self-dual
equation (\ref{eqAB}) it is clear that in the abelian theory the 
bosonic current
conservation is valid as well,
\begin{equation}
\partial_{\mu}j^{\mu}=0.
\label{eqconb}
\end{equation}
This makes the mapping between the currents $J_\mu$ and  $j_\mu$ 
unambiguous. It is important to note that $j_\mu$ is a topological
current, meaning that its conservation is assured by construction.

The Hamiltonian in the static limit simply reduces to the Lagrangian with
a negative sign,
\begin{equation}
H_B=-L_B.
\label{eqhl}
\end{equation}

In the next section, we will show that the abelian bosonised model
has a local gauge invariance. This gauge symmetry along with the 
set of time independent equations of motion helps us to solve the abelian
model completely.
\vskip 1cm

\noindent{\bf III. Abelian theory}
\vskip 0.5cm
In the abelian case,
one can replace the covariant derivatives by simple derivatives and
(\ref {eqAB}) is reduced to
\begin{equation}
2\sigma\epsilon^{\nu\mu\lambda}\partial_\mu B_\lambda+\gamma B^{\nu}=0,
\label{eqABab}
\end{equation}
We are interested in the behaviour of the matter density $B_0$ in the
static limit. Hence all time derivatives are dropped. The above equation
is broken up in component form,
$$\gamma B_0+2\sigma B_{12}=0$$
\begin{equation}
-\gamma B_1+2\sigma\partial_2B_0=0,
~~\gamma B_2+2\sigma\partial_1B_0=0.
\label{eqcur}
\end{equation}
From the last two equations we get,
$$\gamma B_{12}=2\sigma\nabla^2B_0=-{{\gamma^2}\over{2\sigma}}B_0.$$
Combining with the first of (\ref{eqcur}), it is found that
$B_0$ satisfies the time independent Helmholtz equation
\begin{equation}
\nabla^2B_0+({{\gamma}\over{2\sigma}})^2B_0=0.
\label{eqhel}
\end{equation}
In fact the above equation is true for $B_\mu$.

Now we show the gauge invariance in the model. The Lagrangian is
\begin{equation}
L_B=-{a\over 4}C_{\mu\nu}C^{\mu\nu}-{1\over {2g}}B_\mu B^{\mu}
+{\alpha\over 2}\epsilon^{\mu\nu\lambda}C_\mu C_{\nu\lambda}.
\label{eqlab}
\end{equation}
Rewriting 
$$C_\mu=B_\mu+\gamma A_\mu+\sigma K_\mu=B_\mu+\bar{A}_\mu $$
the field tensor breaks in to two decoupled parts,
$$C_{\mu\nu}=B_{\mu\nu}+\bar{A}_{\mu\nu}.$$
In terms of these redefinitions, (\ref{eqlab}) becomes
\begin{equation}
L_B=-{a\over 4}(B+\bar{A})_{\mu\nu}(B+\bar{A})^{\mu\nu}
-{1\over {2g}}B_\mu B^{\mu}
+{\alpha\over 2}\epsilon^{\mu\nu\lambda}(B+\bar{A})_\mu 
(B+\bar{A})_{\nu\lambda}.
\label{eqlab1}
\end{equation}
Clearly the action is invariant up to a total derivative under the local
gauge transformation,
$$\bar{A}_\mu\rightarrow\bar{A}_\mu+\partial_\mu\phi $$
where $\phi $ is some arbitrary function.

This allows us to choose a gauge
\begin{equation}
\bar{A}_0\equiv \gamma A_0+2\sigma A_{12}=0.
\label{eqgauge}
\end{equation}
which makes $\bar{A}_{0i}=0$ and $B_{0i}=-\partial_iB_0$ in the static
case. Using this gauge and static expressions, we simplify the 
components related to $K_\mu $ field,
$$K_\mu=2\epsilon_{\mu\nu\lambda}\partial^\nu A^\lambda,$$
$$K_0=2A_{12},~~K_1=-2\partial_2 A_0,~~K_2=2\partial_1 A_0,$$
$$K^{i0}=2\partial^iA_{12},~~K_{12}=-2\nabla^2A_0.$$
Now, from (\ref{eqB}), for $\nu=0$ we get,
$$({{a\gamma^2}\over{4\sigma^2}}+{{\alpha\gamma}\over\sigma}+{1\over g})
B_0+2\sigma(\nabla^2A_0+({\gamma\over{2\sigma}})^2A_0)=0,$$
(where (\ref{eqhel}) has been used),
and for $\nu=1$ and $\nu=2$ we get,
$$-a\partial_2C_{12}+2\alpha\partial_2B_0+{{B_1}\over g}=0,~
-a\partial_1C_{12}+2\alpha\partial_1B_0-{{B_2}\over g}=0,$$
which are combined to give,
$$({\gamma\over{2\sigma}})
({{a\gamma^2}\over{4\sigma^2}}+{{\alpha\gamma}\over\sigma}+{1\over g})
B_0-2\sigma a\nabla^2(\nabla^2A_0+({\gamma\over{2\sigma}})^2A_0)=0.$$
From the above set of equations, we finally obtain an equation involving
$A_0$ only,
\begin{equation}
\nabla^2(\nabla^2A_0+({\gamma\over{2\sigma}})^2A_0)+
{\gamma\over{2a\sigma}}(\nabla^2A_0+({\gamma\over{2\sigma}})^2A_0)=0.
\label{eqA0}
\end{equation}
Note that for small $a$ we have approximately
$$\nabla^2A_0+({\gamma\over{2\sigma}})^2A_0=0,$$
which is identical to (\ref{eqhel}).

We now consider two special cases:
{\bf (i)} $ \gamma=0,~~\sigma=0$, the Thirring model \cite{th} and
{\bf (ii)} $g=0,~~\sigma=0$, the Deser Redlich model \cite{dr}.
Note that the third option, i.e., bosonisation of the free fermion theory
with 
 $\gamma=\sigma=g=0$ is not permissible in this scheme as $g^{-1}$ is
present.

In Case {\bf (i)},the set of equations of motion in 
(\ref{eqA},\ref{eqB},\ref{eqAB}) now reduce to the single equation.
\begin{equation}
a\partial_\mu B^{\mu\nu}+\alpha\epsilon^{\nu\mu\lambda}B_{\mu\lambda}
-{{B^\nu}\over g}=0.
\label{eqtthb}
\end{equation}
Breaking it into components, we end up with the equations,
\begin{equation}
2\alpha B_{12}+a\nabla^2 B_0-{{B_0}\over g}=0;~~{{B_{12}}\over g}-a\nabla^2
B_{12}+2\alpha\nabla^2B_0=0.
\label{eqb12}
\end{equation}
This reproduces a static equation of motion involving only $B_0$,
\begin{equation}
a^2(\nabla^2)^2B_0-({{2a}\over g}-4\alpha^2)\nabla^2B_0
+{{B_0}\over{g^2}}=0.
\label{eqthb}
\end{equation}
Rewriting the above equation in the form, where $a^2$ has been dropped,
$$\nabla^2B_0+{1\over{({2g\alpha})^2}}(1-{a\over{2g\alpha^2}})^{-1}B_0=0,$$
we make an expansion in $a$,
$$\nabla^2B_0+{1\over{({2g\alpha})^2}}(1+{a\over{2g\alpha^2}}+...)B_0=0.$$
Note that the $B_0$ mass term is renormalised by fermion mass corrections.
With $C_\mu=B_\mu $, the bosonised
Lagrangian and current reduce to the well known forms,
$$
L_B=-{a\over 4}B_{\mu\nu}B^{\mu\nu }-{1\over {2g}}B_\mu B^{\mu }
+\alpha\epsilon^{\mu\nu\lambda}B_\mu\partial_\nu B_\lambda .$$
$$
X^\nu =a\partial_\mu B^{\mu\nu}
+\alpha\epsilon^{\nu\mu\lambda}B_{\mu\lambda}.$$

In Case {\bf (ii)}, for $\sigma=0$,
 $C_\mu=B_\mu+\gamma A_\mu$ and
the Lagrangian becomes,
\begin{equation}
L_{DR}=-{a\over 4}(B+\gamma A)_{\mu\nu}(B+\gamma A)^{\mu\nu }
-{m\over {6\pi}}B_\mu B^{\mu }
+\alpha\epsilon^{\mu\nu\lambda}(B+\gamma A)_\mu
\partial_\nu (B+\gamma A)_\lambda .
\label{eqdr}
\end{equation}
The above Lagrangian breaks up into two pieces, a $B_\mu $ independent one,
\begin{equation}
L_{DR}(A)=-{{a\gamma^2}\over 4}A_{\mu\nu}A^{\mu\nu }
+\alpha\gamma\epsilon^{\mu\nu\lambda}A_\mu
\partial_\nu A_\lambda, 
\label{eqdra}
\end{equation}
and a $B_\mu $ dependent one,
$$
L_{DR}(A,B)=-{a\over 4}B_{\mu\nu}B^{\mu\nu }
+\alpha\epsilon^{\mu\nu\lambda}B_\mu
\partial_\nu B_\lambda $$
\begin{equation}
-{1\over{2g}}B_\mu B^\mu-{{a\gamma}\over 2}B_{\mu\nu}
A^{\mu\nu}+2\alpha\gamma\epsilon^{\mu\nu\lambda} B_\mu\partial_\nu A_\lambda.
\label{eqdrab}
\end{equation}
Rewriting the latter equation in the following form,
\begin{equation}
L_{DR}(A,B)=B_\mu P^{\mu\nu}B_\nu+B_\mu Q^\mu,
\label{eqbpq}
\end{equation}
where,
$$P^{\mu\nu}=-{1\over{2g}}g^{\mu\nu}+{a\over 2}(g^{\mu\nu}\partial^\mu
\partial_\mu-\partial^\mu\partial^\nu)-\alpha\epsilon^{\mu\nu\lambda}
\partial_\lambda,$$
$$Q^\mu=\gamma(a(g^{\mu\nu}\partial^\mu
\partial_\mu-\partial^\mu\partial^\nu)-2\alpha\epsilon^{\mu\nu\lambda}
\partial_\lambda)A_\nu.$$
Performing the gaussian integration for $B_\mu $ leads to the formal
result,
\begin{equation}
L_{DR}(A,B)\approx -{1\over 4}Q^\mu (P^{-1})_{\mu\nu}Q^\nu.
\label{eqbint}
\end{equation}
At present we are only interested in getting local terms with smaller
number of derivatives, and hence we take the inverse of $P_{\mu\nu}$ as simply
$$(P^{-1})_{\mu\nu}\approx -2g g_{\mu\nu}.$$
No gauge fixing is cosidered so far.
Substituting this back in (\ref{eqbint}) yields,
$$
L_{DR}(A,B)={{g\gamma^2}\over 2}
[(a(g^{\mu\nu}\partial^\lambda
\partial_\lambda-\partial^\mu\partial^\nu)-2\alpha\epsilon^{\mu\nu\lambda}
\partial_\lambda)A_\nu]$$
\begin{equation}
[(a(g^{\mu\eta}\partial^\rho
\partial_\rho-\partial^\mu\partial^\eta)-2\alpha\epsilon^{\mu\eta\rho}
\partial_\rho)A_\eta].
\label{eqbint1}
\end{equation}
Keeping in mind the condition of lowest number of derivatives, we take
only the following contribution in the effective action,
\begin{equation}
L_{DR}(A,B)=g(\alpha\gamma)^2 A_{\mu\nu}A^{\mu\nu}=g({\gamma\over{4\pi}})^2
{{6\pi}\over m}A_{\mu\nu}A^{\mu\nu}.
\label{eqbeff}
\end{equation}
Thus we notice that in this order the 
coefficient of the Maxwell term in $A_\mu $ gets modified whereas the 
Chern-Simons term in $A_\mu $ remains unaltered. The final form of the action
to the order stated is,
\begin{equation}
L_{DR}=-\gamma^2({g\over{\pi^2}}-{a\over 4})A_{\mu\nu}A^{\mu\nu }
+\alpha\gamma\epsilon^{\mu\nu\lambda}A_\mu
\partial_\nu A_\lambda, 
\label{eqdrt}
\end{equation}
This is exactly the model studied in \cite{dr}, if following \cite{dr}
 the Thirring 
coupling $g$ is taken to be proportional to $m^{-1}$.

Let us now discuss the effects of higher order (in $m^{-1}$)
 Seelay terms in the 
fermion determinant in the full theory. Considering the next Seelay term
our Lagrangian will be,
\begin{equation}
{\bar L}_B=L_B+y\epsilon^{\mu\nu\lambda}C_\mu \partial^\rho \partial_\rho
\partial_\nu C_\lambda,
\label{eqs3}
\end{equation}
where $L_B$ is given in (\ref{eqlab}) and $y$ is of order $O(m^{-2})$. 
Clearly
the equations of motion in (\ref{eqA}), (\ref{eqB}) and (\ref{eqAB}) will
remain unchanged structurally, with $X_\nu$ changing to ${\bar X}_\nu$,
\begin{equation}
{\bar X}^\nu=X^\nu+y\epsilon^{\nu\mu\lambda}\partial^\rho\partial_\rho 
C_{\mu\lambda}. 
\label{eqxx}
\end{equation}
However, in the full theory this will not change the behaviour of
$B_0$. On the other hand, in the pure Thirring model, the $B_\mu$-equation
in (\ref{eqAB}) is modified to,
\begin{equation}
a\partial_\mu B^{\mu\nu}+\epsilon^{\nu\mu\lambda}(\alpha+y\partial^2)
B_{\mu\lambda}-{{B^\nu}\over g}=0.
\label{eqBB}
\end{equation}
The resulting time-independent equations are,
$${{B_{12}}\over g}-a\nabla^2B_{12}+2(\alpha+y\nabla^2)\nabla^2B_0=0,$$
$$a\nabla^2B_0-{{B_0}\over g}+2(\alpha+y\nabla^2)B_{12}=0.$$
Neglecting terms of $O(ay)$ we get,
$$B_{12}={1\over{2\alpha}}(1+{y\over\alpha}\nabla^2)^{-1}({{B_0}\over g}
-a\nabla^2B_0)$$
$$\approx {1\over{2\alpha}}({{B_0}\over g}-(a-{y\over{\alpha g}})\nabla^2
B_0).$$
Hence the $B_0$-equation becomes,
\begin{equation}
2\alpha a^2(\nabla^2)^2B_0+(4\alpha^2-{{2a}\over g}+{y\over{\alpha g^2}})
\nabla^2 B_0+{{B_0}\over {g^2}}=0.
\label{eqbb}
\end{equation}
Comparing with (\ref{eqthb}) we note that now the first term can not be
ignored. The next Seelay term, 
$(\epsilon^{\mu\nu\lambda}B_\mu B_{\nu\lambda})^2$ obviously causes more
complications in the pure Thirring model and obtaining an equation involving
$B_0$ only is non-trivial. In the full theory, the generic feature is that
these type of changes leaves the $B_0$-equation intact.

\vskip 1cm
{\bf IV.    Non-abelian Theory:}
\vskip 0.5cm
As has been emphasised before \cite{bbg}, the results are far more
complicated in
the non-abelian scenario. For arbitrary non-abelian gauge fields $A_\mu^a$,
identification between the fermi and bose currents is problematic.
From the equations of motion given in
(\ref{eqA}), (\ref{eqB}) and (\ref{eqAB}), the following covariant
conservation equation emerges,
\begin{equation}
(D_\mu^{(C)}j^\mu)^a=0.
\label{eqnonab})
\end{equation}
But this is different from (\ref{eqferc}). Also there is no local
gauge invariance in the nonabelian bosonised version 
due to the nature of cross terms between $B_\mu^a$ and $\bar{A}^a_\mu$
present in the theory.

 However, these problems can be completely removed
 for a restricted class of gauge fields, formerly used in
\cite{jp}, that are proportional to the generators of the Cartan
subalgebra only,
\begin{equation}
[h^\alpha,h^\beta]=0,~~A_-=\sum_{\alpha=1}^{r}A_\alpha h^\alpha,~~
A_+=-\sum_{\alpha=1}^{r}A^*_\alpha h^\alpha,
\label{eqAcar}
\end{equation}
where 
$$A_{\pm}=A_1\pm iA_2.$$ 
In the fermion problem 
\cite{jp} it was assumed that the fermion fields $\psi$ are 
proportional to the ladder generators  $e^\alpha$ only,
\begin{equation}
\psi=\psi_\alpha e^{\alpha},
\label{eqpsi}
\end{equation}
where,
$$[e^\alpha,e^{-\beta}]=\delta_{\alpha\beta}h^\alpha,~~[h^\alpha,e^\beta]
=K_{\beta\alpha}e^\beta,~~[h^\alpha,e^{-\beta}]=-K_{\beta\alpha}e^{-\beta}.$$
Note that $(e^\beta)^+=-e^{-\beta},~(h^\beta)^+=h^\beta $ and 
the Cartan Matrix $K_{\alpha\beta}$ is real and for $SU(N)$ symmetric. 
Thus in the fermion model the charge is 
$$J_0\approx[\psi^+,\psi]\approx h^\alpha $$
this shows
that the charge is also in the Cartan subalgebra. This
ansatz prompts us to restrict $B_\mu^a$ in the Cartan subalgebra. But with 
$A_\mu^a$ already in the Cartan
subalgebra the entire system is reduced 
to essentially an abelian one, with just a non-interacting index tagged
along each of the fields, reminding us of the nonabelian nature. Hence,
in the lowest order of inverse fermion mass, we
get a number of decoupled static
Helmoltz equations for the non-abelian charge
$B_0^a$,
\begin{equation}
\nabla^2B_0^a+({{\gamma}\over{2\sigma}})^2B_0^a=0.
\label{eqheln}
\end{equation}
\vskip 1cm
{\bf V.  Conclusion}
\vskip .5cm
As an application of 2+1-dimensional bosonisation, we have studied
thoroughly the non-minimally gauged massive Thirring model. Computing
the fermion determinant up to first order in inverse fermion mass, the
charge in the abelian model is shown to obey the (static) massive
Helmholtz equation. Special cases leading to known results in the Thirring
and Deser-Redlich models are derived. For abelian gauge group,
effects of higher order terms are also
discussed. In case of non-abelian gauge fields, a restricted class of
gauge fields reduces the system to essentially a group of decoulped
abelian ones and the charges behave in an identical fashion to the
abelian one.
\vskip 1cm
{\bf Acknowledgements.}
\vskip .5cm
It is a pleasure to thank Dr. Rabin Banerjee for many helpful discussions.
Also I am grateful to Professor C. K. Majumdar, Director, S. N. Bose
National Centre for Basic Sciences, for allowing me to use the Institute
facilities.

\end{document}